\documentclass[prl, twocolumn, a4paper, superscriptaddress, showpacs, amsmath, amssymb, footinbib, floatfix]{revtex4}

\usepackage[utf8]{inputenc}
\usepackage[english]{babel}
\usepackage{graphicx}
\usepackage{bm}
\usepackage{color}

\begin{document}

\renewcommand{\figurename}{FIG.}

\title{Criticality of relaxation in dislocation systems}

\author{P\'eter Dus\'an Isp\'anovity}
\email{ispanovity@metal.elte.hu}
\homepage{http://dislocation.elte.hu}
\affiliation{Paul Scherrer Institut, CH-5232 Villigen PSI, Switzerland}
\affiliation{Department of Materials Physics, E\"otv\"os University Budapest,
H-1517 Budapest POB 32, Hungary}

\author{Istv\'an Groma}
\affiliation{Department of Materials Physics, E\"otv\"os University Budapest,
H-1517 Budapest POB 32, Hungary}

\author{G\'eza Gy\"orgyi}
\affiliation{Department of Materials Physics, E\"otv\"os University Budapest,
H-1517 Budapest POB 32, Hungary}

\author{P\'eter Szab\'o}
\affiliation{Department of Materials Physics, E\"otv\"os University Budapest,
H-1517 Budapest POB 32, Hungary}

\author{Wolfgang Hoffelner}
\affiliation{Paul Scherrer Institut, CH-5232 Villigen PSI, Switzerland}

\begin{abstract}
Relaxation processes of dislocation systems are studied by two-dimensional dynamical simulations.  In order to capture generic features, three physically different scenarios were studied and power-law decays found for various physical quantities.  Our main finding is that all these are the consequence of the underlying scaling property of the dislocation velocity distribution.  Scaling is found to break down at some cut-off time increasing with system size.  The absence of intrinsic relaxation time indicates that criticality is ubiquitous in all states studied.   These features are reminiscent to glassy systems, and can be attributed to the inherent quenched disorder in the position of the slip planes.
\end{abstract}


\maketitle

When crystalline materials are subjected to large enough stresses they undergo plastic, irreversible deformation caused by the motion of dislocations. As it is well-known, these linear lattice defects interact via long-range ($1/r$ type) stress fields \cite{Hirth1982} playing a crucial role in several complex phenomena related to plasticity, such as the formation of various dislocation patterns during deformation \cite{Zaiser2002} and dislocation avalanches \cite{Csikor2007}.  Other systems with long-range interactions, like gravitating particles \cite{Padmanabhan1990} or non-neutral plasmas \cite{Nicholson1983} have been intensely studied and found to exhibit several unique properties, like power-law relaxation \cite{Campa2009}.  Unlike dislocations these systems are Hamiltonian, still one would expect similar level of complexities.

Another characteristics of dislocation systems is the inherent randomness in the positions of slip planes, wherein individual dislocations glide easily. On the other hand, systems with disorder, like structural and spin glasses, have been the focus of much interest.   They were found to show peculiar dynamical properties \cite{Ilyin2008}, such as slow relaxation, attributed to a wide spectrum of decay times. The fact that dislocation systems contains quenched disorder through the  glide planes raises the analogy with spin glasses. Glassy dynamical behavior has indeed been observed in simulations \cite{Bako2007} and experimentally \cite{Gerbode2010} for dislocation systems, but the phenomenon is still lacking a systematic study.

The interplay of long-range interaction and disorder leads to complex behavior even in two-dimensional (2D) dislocation systems \cite{Laurson2010, Rosti2010, Tsekenis2011}.  The model is a strong simplification over reality, local processes \cite{Gomez2006} were neglected, only justified by the richness of phenomena reported about in this paper.  Slow relaxation was observed in several instances, like under constant external stress, i.e., creep condition.  In this case a dislocation system with single slip exhibits the well-known Andrade-type creep law \cite{Nabarro2006}, with the plastic strain rate $\dot{\gamma}_\text{pl}$ decreasing in time $t$ as $\dot{\gamma}_\text{pl}(t) \sim t^{-2/3}$ until a cut-off time $t_1$ \cite{Miguel2002}. It was also suggested that $t_1$ tends to infinity as some critical stress is approached from below, hinting to an analogy between the yielding transition and conventional phase transitions \cite{Miguel2002, Zaiser2006}. In addition, it was also reported that single slip random 2D dislocation systems at zero external stress relax to an equilibrium state slowly, with a relaxation time increasing with system size \cite{Csikor2009}. Such slow relaxation processes of dislocated crystals have also been observed experimentally \cite{Nabarro2006, Louchet2009}.

In this Letter we focus on the properties of relaxation to equilibrium below the yield stress.  In all arrangements studied we observe power-law decay of various quantities, a feature due to the underlying scaling of the dislocation velocity distribution.   Scaling is always found to cut off at a characteristic time increasing with system size, indicating that critical behavior is not limited to a given threshold stress suggested by earlier investigations \cite{Miguel2002, Laurson2010}, rather,  criticality is present in all states studied.

The system considered is the simplest representation of a dislocated crystal, consisting of parallel straight edge dislocations with parallel slip planes. Thus the problem is simplified to 2D. By denoting the position of the $i$th dislocation by $\bm r_i = (x_i, y_i)$, its Burgers vector by $\bm b_i = s_i(b,0)$ ($s_i = \pm 1$ is the sign of the ``charge''), the equation of motion of a dislocation can be written as \cite{Miguel2002}
\begin{equation}
	\dot{x}_i = s_i \left[ \sum_{j=1;\ j\ne i}^{N} \!\!\! s_j \tau_\text{ind}(\bm
r_i - \bm r_j) + \tau_\text{ext}(\bm r_i) \right]\!\!;\ \dot{y}_i = 0.
\label{eqn:eq_mot}
\end{equation}
Here $\tau_\text{ind}(\bm r) = \cos (\varphi) \cos (2\varphi)r^{-1}$ is the shear stress field generated by an individual dislocation, $\tau_\text{ext}$ denotes the external shear stress, and $N$ is the total number of the constituent dislocations. We note that the different physical parameters are absorbed in the length-, time-, and stress-scales, as we measure them in $\rho^{-0.5}$, $(\rho M Gb^2)^{-1}$, and $Gb\rho^{0.5}$ units, respectively, where $\rho$ is the dislocation density, $M$ is the dislocation mobility, and $G$ is an elastic constant \cite{Csikor2009}. So, in the rest of this Letter, only these dimensionless units are used.

Let us highlight the most important physical features of this conceptually simple model. Firstly, the pair interaction is of long-range character, since it decays with $1/r$, moreover, it exhibits a complicated angular dependence with zero average. Secondly, the equation of motion (\ref{eqn:eq_mot}) is first order, which is due to the commonly assumed overdamped nature of dislocation motion, thus the system is strongly dissipative, i.e., non-Hamiltonian. Thirdly, the motion is constrained, since dislocations can only move parallel to the $x$ axis. This has the important consequence that if the initial $y$ coordinates are chosen randomly then this will represent quenched disorder. As a result, the system never completely forgets its initial state, that is, it does not collapse into a global energy minimum, rather gets trapped into a local minimum. Thus, the ground state of the system is frustrated leading to a glassy-like dynamics \cite{Bako2007}.

\begin{figure}[!ht]
\begin{center}
\hspace*{-1cm}
\begin{picture}(0,0)
\put(162,-24){\sffamily{(a)}}
\end{picture}
\includegraphics[angle=-90]{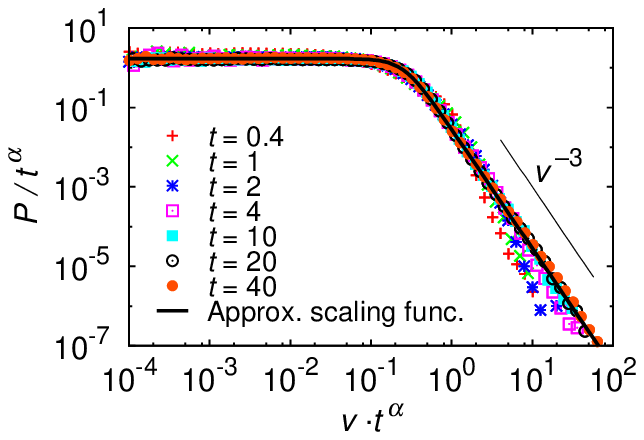} \\
\hspace*{-1cm}
\begin{picture}(0,0)
\put(162,-33){\sffamily{(b)}}
\end{picture}
\includegraphics[angle=-90]{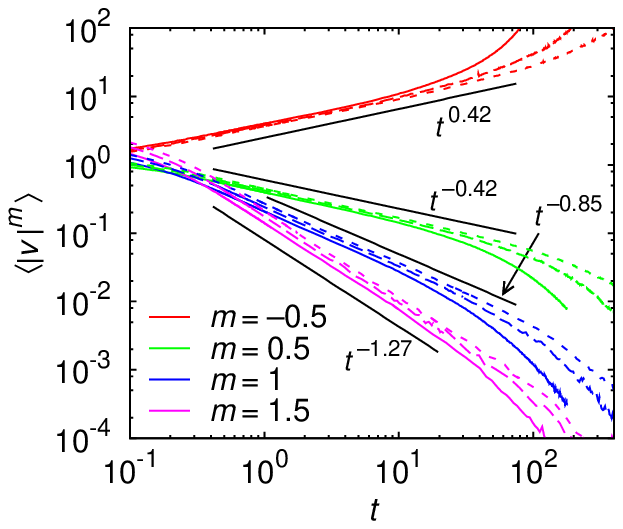}
\end{center}
\caption{(color online) Dynamics during relaxation from a random configuration.
(a) The scaled velocity distributions of dislocations $P_v$ given by Eq.~(\ref{eqn:scaling}) with $\alpha=0.85$ at system size $N=2048$, and the approximated fitted scaling function (see text).
(b) The moments $\langle |v(t)|^m \rangle$ for different exponents $m$ (identified by colors) and dislocation numbers $N$. The solid, dashed, and dotted lines correspond to $N=128$, $512$, and $2048$, respectively.
\label{fig:relax}}
\end{figure}

In the first part of this Letter the relaxation of random dislocation systems is studied. In this scenario an equal number of positive ($s_i=1$) and negative ($s_i=-1$) sign dislocations are placed randomly in a square-shaped area. Then, with periodic boundary conditions, the equations of motion (\ref{eqn:eq_mot}) are solved numerically \cite{Csikor2009} until no considerable dislocation movement is observed (for an example simulation movie see \footnote{\hyphenation{supplemental} See supplementary material at http://link.aps.org/ supplemental/?\ for movies on single relaxation runs and illustration of cut-off times.}). It is noted that due to the periodic boundary conditions, the stress field $\tau_\text{ind}$ is also modified, for details see \cite{Bako2006}. The simulations were repeated with different initial configurations $13\,000$, $300$, and $118$ times for the system sizes of $N=128$, $512$, and $2048$, respectively. During the relaxation of the system, the time dependent probability distribution of the dislocation velocities $P(v,t)$ was determined numerically. (Since there is no external stress, for symmetry reasons, the velocity distribution of positive and negative dislocations are equal and symmetric.) According to Fig.~\ref{fig:relax}(a) a remarkable feature of the velocity distribution is that it decays to zero as $v^{-3}$ (for a theoretical explanation see \cite{Ispanovity2010}). In addition, $P(v,t)$ tends to a Dirac-delta function (corresponding to the equilibrium state) and between an initial $t_0$ and a cut-off time $t_1$ this is described with the scaling law
\begin{equation}
	P(v, t) = t^\alpha f(t^\alpha v).
\label{eqn:scaling}
\end{equation}
The exponent $\alpha$ was found to be $\alpha = 0.85(2)$ and the scaling function $f$ can be well approximated by the form $f(x) \approx A/(Bx^{3}+1)$. Note, that the curves plotted are always results of averaging over the statistical ensemble.

In order to investigate this scaling behavior in more detail, the $m$th moment $\langle |v(t)|^m \rangle$ of the absolute velocity was also determined for different $m$ values. According to Eq.~(\ref{eqn:scaling})
\begin{equation}
	\langle |v(t)|^m \rangle = \int |v|^m P(v, t) d v = C_m t^{-m \alpha},
\label{eqn:moments}
\end{equation}
where $C_m$ is a constant. (Because of the asymptotic properties of the scaling function $f$, the integral is finite only for $-1<m<2$.) Figure \ref{fig:relax}(b) shows the measured $\langle |v(t)|^m \rangle$ curves for different $m$ values and system sizes. The fitted exponents are in agreement with Eq.~(\ref{eqn:moments}). (With the same simulation setup the scaling regime and the value of $\alpha$ has been already reported in \cite{Csikor2009}.)

As seen in Fig.~\ref{fig:relax}(b) the scaling region is bounded from both sides. It starts at a fixed $t_0\approx0.4$ and lasts till a cut-off time $t_1$ increasing with increasing linear system size $L=\sqrt{N}$. This is in perfect accordance to what is usually found in systems with long-range interactions. The actual $t_1$ versus $L$ relation is analyzed below.

\begin{figure}[!t]
\begin{center}
\includegraphics[angle=-90]{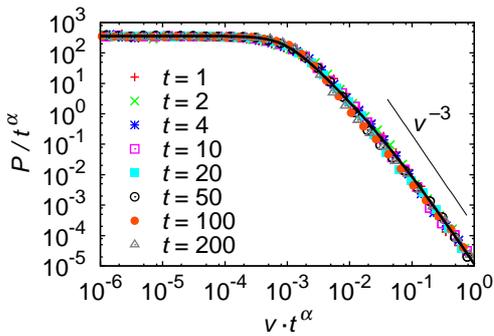}
\end{center}
\caption{(color online) Scaling of the velocity distribution after adding an extra fixed dislocation to a relaxed system with $\alpha = 0.34$ at $N=2048$. The corresponding scaling function (solid line) is $f(x)\approx A/(Bx^3+Cx^2+1)$.
\label{fig:oneextra_scaling}}
\end{figure}

In the second simulation scenario, an extra dislocation with fixed position is introduced to an already relaxed configuration \footnotemark[\value{footnote}]. This can be considered as the prototype of an external perturbation. (The equilibrium dislocation configuration generated by the extra dislocation, the analogue of Debye-screening, was analyzed in \cite{Groma2006, Ispanovity2008}.) Due to the extra force field, the system evolves to a new equilibrium state. As Fig.~\ref{fig:oneextra_scaling} shows $P(v,t)$ obeys again the same scaling law given by Eq.~(\ref{eqn:scaling}), now with $\alpha = 0.34(4)$. Consistently, the evolution of different velocity moments obey Eq.~(\ref{eqn:moments}), and thus confirms scaling.

\begin{figure}[!t]
\begin{center}
\hspace*{-1cm}
\begin{picture}(0,0)
\put(162,-24){\sffamily{(a)}}
\end{picture}
\includegraphics[angle=-90]{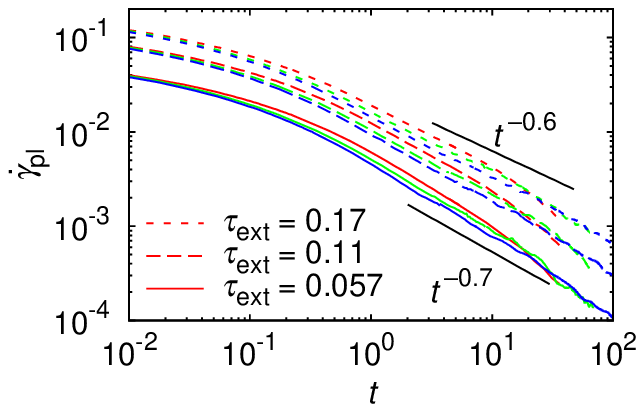} \\
\begin{picture}(0,0)
\put(105,-24){\sffamily{(b)}}
\put(195,-24){\sffamily{(c)}}
\end{picture}
\includegraphics[angle=-90]{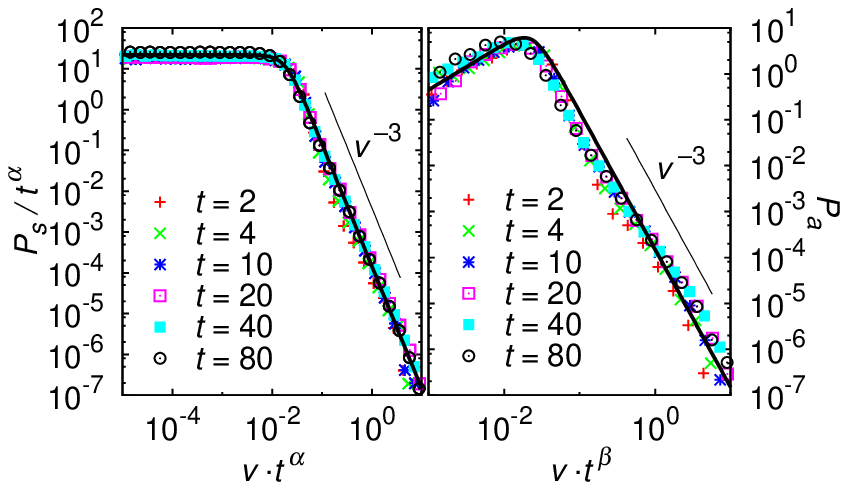} \\
\end{center}
\caption{(color online) Relaxation due to a constant external shear stress $\tau_\text{ext}$ below the yield stress.
(a) The mean plastic strain rate $\dot{\gamma}_\text{pl}$ versus time $t$ curves for different $\tau_\text{ext}$ and system sizes $N$. Curves for increasing size (red, green, and blue) correspond to $N=128$, $512$, and $2048$, respectively.  The plot centers on the power-decaying regime, not resolving the cut-off times.
(b) The symmetric part of the velocity distribution $P_s$ at different times scaled with Eq.~(\ref{eqn:pull_scaling}) with $\alpha=0.32$ at $N=2048$ and $\tau_\text{ext} = 0.11$. The approximate scaling function (solid line) is $f(x) \approx A/(Bx^3 + 1)$.
(c) The antisymmetric part of the velocity distribution $P_a$ at different times scaled with Eq.~(\ref{eqn:pull_scaling}) with $\beta=0.32$ at $N=2048$ and $\tau_\text{ext} = 0.11$. The approximate scaling function (solid line) is $g(x) \approx C/(Dx^3 + x^{-1})$.
\label{fig:pull}}
\end{figure}

In the third scenario an external constant shear stress $\tau_\text{ext}$ is turned on to the relaxed system \footnotemark[\value{footnote}] (creep experiment first studied by Miguel et al.~\cite{Miguel2002, Rosti2010}). If $\tau_\text{ext}$ is smaller than a certain yield stress, the induced mean plastic strain rate $\dot{\gamma}_\text{pl}(t)$ decreases to zero. Figure~\ref{fig:pull}(a) shows the evolution of $\dot{\gamma}_\text{pl}$ at different external stresses $\tau_\text{ext}$ and system sizes $N$. As seen $\dot{\gamma}_\text{pl}$ also has a power-law regime with exponent slightly decreasing with increasing $\tau_\text{ext}$. Like for the relaxation without external stress, the power decay is cut off at some time  $t_1$, not shown in Fig.~\ref{fig:pull}(a) for clarity. We again find that $t_1$ is increasing with system size, for more details see \footnotemark[\value{footnote}].  It is, however, practically independent from $\tau_\text{ext}$, a conclusion different from what was drawn by Miguel et al.~\cite{Miguel2002,Rosti2010}. Namely, they found that at a critical stress level $\dot{\gamma}_\text{pl}(t)$ becomes pure power-law without a cut-off time, whereas below it a stress-dependent cut-off exists.  According to our investigations for low stresses the cut-off $t_1\!\propto \! \sqrt{N}$ is mainly due to finite size.

We now turn to the velocity distribution of positive  and negative dislocations,  denoted by $P^+$ and $P^-$, respectively.  For symmetry reasons the distribution always fulfills $P^+(v, t) = P^-(-v, t)$, so it is enough to investigate $P^+$, denoted hereafter by $P$. We separate it into a symmetric and an antisymmetric part as $P = P_s + P_a$ (note that in the absence of external stress $P_a=0$, hence $P^+=P^-$).  Then in the definition of the velocity moments [Eq.~(\ref{eqn:moments})] $P$ has to be replaced by $P_s$, and $\dot{\gamma}_\text{pl}$ is
\begin{equation}
	\dot{\gamma}_\text{pl}(t) = \bigg\langle \sum_{i=1}^{N} s_i v_i \bigg\rangle = \int v P_a(v, t) d v.
\label{eqn:gamma_dot_def}
\end{equation}
Figure~\ref{fig:pull}(b) and (c) show the scaling of $P_s$ and $P_a$, respectively, for $\tau_\text{ext}=0.11$ with
\begin{equation}
	P_s(v,t) = t^\alpha f(t^\alpha v) \quad \text{ and } \quad P_a(v, t) = g(t^\beta v),
\label{eqn:pull_scaling}
\end{equation}
at $\alpha = 0.32(2)$ and $\beta = 0.32(3)$, numerically indistinguishable in this case. Note, that no power prefactor was found in the scaling form of $P_a$ above. From Eqs.~(\ref{eqn:gamma_dot_def},\ref{eqn:pull_scaling}) $\dot{\gamma}_\text{pl}(t) = C t^{-2\beta}$, with an appropriate $C$ constant. This is in complete agreement with the time evolution of $\dot{\gamma}_\text{pl}$ plotted in Fig.~\ref{fig:pull}(a). For different $\tau_\text{ext}$ values, the same scaling is found with slightly different exponents. We mention that the plastic strain $\gamma_\text{pl}(t)$ has the exponent $1-2\beta$, this varies in the range $0.3-0.4$, and is in accordance with the well-known power 1/3 of the Andrade creep \cite{Nabarro2006, Miguel2002}.

To summarize the above findings, the scaling formulas Eq.~(\ref{eqn:pull_scaling}) seem to be generally valid. The exponents vary, however, a feature we attribute to the difference in the initial conditions and external stress field, the only properties distinguishing the scenarios we considered.  Thus, the exponent is not a universal, inherent characteristics of the dislocation system determined only by the interactions and dimensionality.

\begin{figure}[!t]
\begin{center}
\hspace*{-1cm}
\includegraphics[angle=-90]{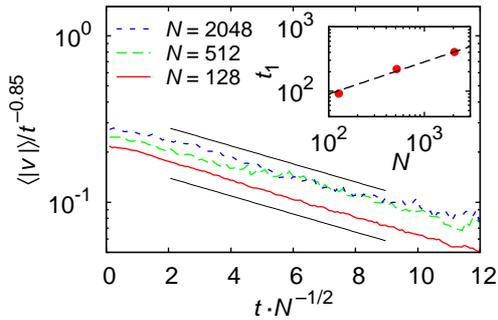}
\end{center}
\caption{(color online) Rescaled mean absolute velocity $t^{\alpha} \langle|v(t)|\rangle$ as a function of scaled time $t/\sqrt{N}$ at different system sizes. Straight line on the semilog plot shows the exponential component, defining a relaxation time $t_1$ for each $N$. Inset shows the obtained $t_1$ values and the $t_1 \sim \sqrt{N}$ fit.
\label{fig:cutoff}}
\end{figure}

The last issue to be elucidated is the system size dependence of the cut-off time $t_1$. In order to determine $t_1$, $t^{\alpha}\langle |v(t)|\rangle$ is plotted with a semilogarithmic scale in Fig.~\ref{fig:cutoff} for the first simulation scenario. As seen, the curves obtained can be well fitted by straight lines, so the time evolution of $\langle|v|\rangle$ can be described with the form $\langle|v(t)|\rangle = Ct^{-\alpha}\exp(-t/t_1)$. Thus the cut-off time $t_1$ is a simple relaxation time.
In addition, in Fig.~\ref{fig:cutoff} the time is rescaled with $L=\sqrt{N}$ and with this the curves corresponding to different system sizes are parallel, indicating that $t_1$ is proportional to the system size $L$. Similar behavior is found for the different order velocity moments introduced above. These results indicate that the relaxation of $\langle|v(t)|\rangle$ can be described with the rate equation
\begin{equation}
\frac{d \langle|v(t)|\rangle} {dt}=-\left(\frac{\alpha}{t}+\frac{1}{t_1}
\right)\langle|v(t)|\rangle,
\end{equation}
where $t_1 \propto L$. This can be interpreted as follows: at large enough time ($t\gg t_1$) during the relaxation of a dislocation system with finite size it always gets close to its energy minimum and enters into an exponential relaxation regime. With increasing system size, however, the energy landscape becomes more and more complex resulting in a longer time to reach the exponential relaxation regime. This is in agreement with the recent results of Laurson et al.~\cite{Laurson2010}, who found that the dynamics slow down dramatically as the yield stress is approached from above. We would also like to highlight that although the time exponent $\alpha$ varies with conditions of the relaxation, the size-dependence of the cut-off time $t_1$ is in all scenarios studied in this Letter close to linear \footnotemark[\value{footnote}].

In summary, power-law relaxation of dislocation systems was observed in different scenarios.  This effect may be attributed to the quenched random positions of the slip axes and the complex nature of the interactions.   Remarkably, the scaling of the time-dependent velocity distribution goes with different exponents depending on the physical setup.  Scaling is cut off  due to the finite size, so the system does not possess any inherent time scale. The dislocation system is, therefore, found to behave like a critical one  in all cases considered, strongly resembling glassy systems.

\begin{acknowledgments}
P.D.I. thanks P.~Derlet for stimulating discussions. Financial supports of the Hungarian Scientific Research Fund (OTKA) under Contract No.\ K 67778 and The European Union and the European Social Fund under the Grant Agreement No.~T{\' A}MOP-4.2.1/B-09/1/KMR-2010-0003 are gratefully acknowledged.
\end{acknowledgments}

\end{document}